# PERFORMANCE IMPROVEMENT FOR PAPR REDUCTION IN LTE DOWNLINK SYSTEM WITH ELLIPTIC FILTERING

Snikdho Sworov Haque[1], Md. Munjure Mowla[2]

[1,2] Department of Electronics and Telecommunication Engineering, Rajshahi University of Engineering & Technology, Rajshahi, Bangladesh

## ABSTRACT

*This paper is concerned with the performance improvement of PAPR reduction of orthogonal frequency division multiplexing (OFDM) signal using amplitude clipping & filtering based design. Note that OFDM is one of the well adept multi-carrier multiplexing transmission scheme which has been implemented in long term evolution (LTE) downlink. Nonetheless peak to average power ratio (PAPR) is the more rattling problem with OFDM, consequently in this paper a reduction procedure of the PAPR by using amplitude clipping and filtering is proposed. Here we used IIR bandpass elliptic filter after amplitude clipping to reduce the PAPR. The performance of the system in terms of bit error rate (BER) is also canvased as a new filter based clipping method. Our results show that the proposed methodology of clipping method with the IIR elliptic band pass filter significantly reduces the PAPR value.*

## 1. INTRODUCTION

Moving towards the 4th generation (4G) technologies, long term evolution (LTE) is a system which is adopted by 3GPP and designed to get more capacity and speed within wireless networks data delivery at a comparatively lower cost. Voice communication & usage of data has grown very fast in modern era and users wanted using the network connection like broadband. Most of the transmission systems have to face much degradation in unfriendly environment such as multipath, large noise, interference, attenuation, nonlinearities, time variance but at a time they have to maintain constraints like crest factor & limitation of transmit power [1]. These conditions are achieved by multi-carrier modulation and among them orthogonal frequency division multiplexing (OFDM) is most efficient. So LTE has adopted this multicarrier OFDM as its downlink spectrum system. But it comes with large envelope fluctuation and is quantified as peak to average power ratio (PAPR). This is the prima disadvantage of OFDM transmission. In almost all low-cost situations, the limitation of high PAPR looked over its potential benefits [2]. For operating in a perfectly linear region the operated power should keep below the available power. This is the reason that power amplifier is used at the transmitter side. Lot of algorithms has been developed for the reduction of this PAPR. They have their own advantages and limitations [3].

Umpteen approaches have been developed & implemented to minimize effect of PAPR with the expense of more transmit power, bit error rate, loss in data rate & computational complexity. A system trade-off is definitely needed. Methods as peak windowing, peak cancellation, peak reduction carrier(PRC),envelope scaling, amplitude clipping and filtering, decision-aided

DOI : 10.5121/ijcnc.2015.7104                                                                                                           51

reconstruction, coding, partial transmit sequence (PTS), selective mapping (SLM)[4], clustered OFDM, interleaving, tone reservation (TR), active constellation extension (ACE), pilot symbol assisted modulation, companding technique (CT)[5] , tone injection (TI)[6] have been presented earlier[7]. In case of applying the partial transmit sequence (PTS)[8] and selected mapping (SLM) [9-10] procedure, these two have more complexity than that of ACF technique. If another technique named Tone Reservation (TR) is considered, it also allows the data rate loss along with power increasing. As well as the techniques such as Active Constellation Extension (ACE) and the Tone Injection (TI)[6] having criteria of increasing power will be unexpected in case of power constraint situation. A special technique described in [1] can be used to avoid this issue.

Many methods for clipping and filtering have been applied likewise Least Square, Kaiserwood(LSK), Extra-ripple bandpass, Specific Multiple Independent Approximation Errors(SMIAR), Raised Cosine and some others. In past research works through a linear-phase FIR filter depending on the Parks-McClellan algorithm have been used in the composed filtering [11]. Existing method [7] uses the FIR based band pass filter in composite filtering and found the remarkable in case of PAPR reduction.

As a consequence in this paper we used the IIR band pass elliptic filter for designing the filtering technique which significantly reduces PAPR than other techniques.

## 2. THEORETICAL BACKGROUND OF OFDM AND PAPR

OFDM dissevers the frequency spectrum into sub-bands midget enough that the channel effectiveness is flat over a given sub-band. Then a classical In phase Quadrature modulation like BPSK, QPSK, M-QAM is done over the sub-band. If it is properly designed then channel's fast changing effects disappear like they are appearing during the transmission of single symbol and are treated as flat fading at receiver end. A large number of prose cutely spaced orthogonal subcarriers are used for carrying data. The data is carved up into several parallel data streams or channels. Whole data rate is maintained similar as the conventional single carrier modulation scheme with the same bandwidth. Orthogonal Frequency Division Multiplexing (OFDM) is the promising procedure for achieving high data rate and combating with multipath fading in wireless communications.[7]

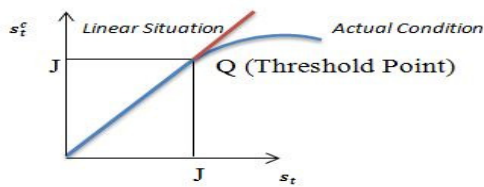

Figure 1. Clipping Function [7]

Linear power amplifiers are being used in the transmitter so that the Q-point has to be in the linear region. Because of the high PAPR the Q-point moves to the saturation region hence the clipping of signal peaks occur and generate in-band and out-off band distortion. For keeping the Q-point in the linear region dynamic range of the power amplifier must be increased which can reduce its efficiency and enhances the cost. So a trade-off exists between nonlinearity and efficiency. And also with the increasing of this dynamic range, cost of power amplifier also increases in parallel, as telecommunication engineer my objective should be to quantize this PAPR which is defined as the ratio between the maximum power and the average power for the envelope of a baseband complex signal $\tilde{s}(t)$ i.e.[1]

$$\text{PAPR} = \frac{\text{Highest Peak Power}}{\text{Total Average Power}}$$

$$\text{PAPR}\{\tilde{s}(t)\} = \frac{\max |\tilde{s}(t)|^2}{E |\tilde{s}(t)|^2} \qquad (1)$$

Also we can write this PAPR equation for the complex pass band signal s(t) as,

$$\text{PAPR}\{s(t)\} = \frac{\max |s(t)|^2}{E |s(t)|^2} \qquad (2)$$

Equation (2) can be re written as,

$$\text{PAPR}\{s(t)\} = \frac{\max_{0 \le t \le UT} |s(t)^2|}{P_{avg}}$$
$$= \frac{\max_{0 \le t \le UT} |s(t)^2|}{\frac{N}{T}\int_0^{UT} |s(t)^2| \, dt} \qquad (3)$$

Where, $P_{avg}$ is the average power and it can be computed in the frequency domain as Inverse Fast Fourier Transform (IFFT) is a scaled unitary transformation. For estimated PAPR of continuous time varying OFDM signals, the OFDM signals samples can be obtained by V times oversampling. V times oversampled time domain samples are UV point IFFT of the data block along with (V-1)U zero-padding. However, the oversampled IFFT output can be expressed as,

$$X[n] = \frac{1}{\sqrt{U}} \left( \sum_{k=0}^{U-1} X_k \exp\frac{j2\pi nk}{UV} \right) \qquad (4)$$

## 3. AMPLITUDE CLIPPING & PROPOSED METHODOLOGY

For 3GPP LTE downlink system the easiest technique which could be used for PAPR reduction is to clip the signal amplitude and then filtering the signal. To do limit the peak envelope or amplitude of the input signal a threshold value of the amplitude has been made fixed here [11]. Clipping ratio (CR) is defined as,

$$CR = \frac{J}{\sigma} \qquad (5)$$

Where, **J** is the amplitude of the signal and σ is the root mean squared value of the unclipped OFDM signal. Before the D/A conversion, the clipping function is performed in digital time domain and the process is described by the following expression,

$$s_t^c = \begin{cases} s_t & ; \ s_t \le J \\ Je^{j\varphi(s_t)} & ; \ s_t > J \end{cases} \qquad (6)$$

Where, $s_t^c$ is the clipped signal, $s_t$ is the transmitted signal, $J$ is the amplitude and $\emptyset(s_t)$ is the phase of the transmitted signal, $s_t$.

BER is aggravated caused by indicating the second point of limitation [2] which is clipped signal passed through the band pass filter (BPF).

The proposed method is shown in the figure 2. Here, through a simplified block diagram of a PAPR reduction scheme is shown using amplitude clipping and filtering, where N is the number of subcarriers and L is the oversampling factor. The input of the IFFT block is the interpolated signal introducing U(V −1) zeros in the middle range can be expressed as the original signal which is,

$$s'[t] = \begin{cases} s[t] & ; \text{for } 0 \leq t \leq U/2 \text{ and } UV - U/2 < t < UV \\ 0 & ; \text{elsewhere} \end{cases} \quad (7)$$

In this system, the V-times oversampled discrete-time signal can be generated as,

$$s'[m] = \frac{1}{\sqrt{UV}} \left( \sum_{t=0}^{UV-1} S'[t] \exp^{\frac{j2\pi\Delta ft}{UV}} \right) \quad ; m = 0, 1, \ldots\ldots UV\text{-}1 \quad (8)$$

Here after, the modulated carrier frequency $f_c$ to yield a passband signal $s^e[m]$.

If $P$ is the pre-specified clipping level then let, $s_q^e[m]$ denote the clipped version of $s^{e'}[m]$ which is expressed as,

$$s_q^e[m] = \begin{cases} -P & ; \text{ se'}[m] \leq -P \\ \text{se'}[m] & ; |\text{se'}[m]| < P \\ P & ; \text{ se'}[m] \geq P \end{cases} \quad (9)$$

After clipping, the signals are passed through the proposed Composed Filter.

The filter itself consists on a set of FFT-IFFT operations where filtering takes place in frequency domain after the FFT function. The FFT function transforms the clipped signal $s_q^e[m]$ to frequency domain yielding $S_q^e[t]$. The information components of $S_q^e[t]$ then are passed to a high pass filter (HPF) producing $\tilde{S}_q^e[t]$. This filtered signal is passed to the unchanged condition of IFFT block and the out-of-band radiation that fell in the zeros is set back to zero. The IFFT block of the filter transforms the signal to time domain and thus obtain $\tilde{s}_q^e[m]$. Proposed Algorithm for PAPR Reduction is given in figure 3.

International Journal of Computer Networks & Communications (IJCNC) Vol.7, No.1, January 2015

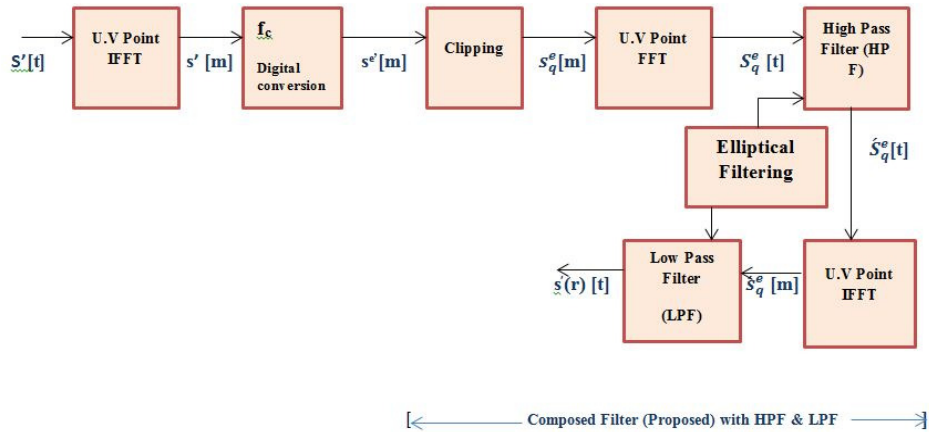

Figure 2. Block Diagram of Proposed Methodology

## 3.1 Elliptical Filtering

An elliptic filter[12] cognized as a signal processing filter with equi-ripple behavior in passband and stopband. The amount of ripple in each band is individually adjustable so none of other filter having equal order can be fast in transition gain between the passband and the stopband for the given values of ripple. Alternatively anyone can minimize the ability to independently adjust the passband and stopband ripple and design a filter. This can be insensitive to component variations.

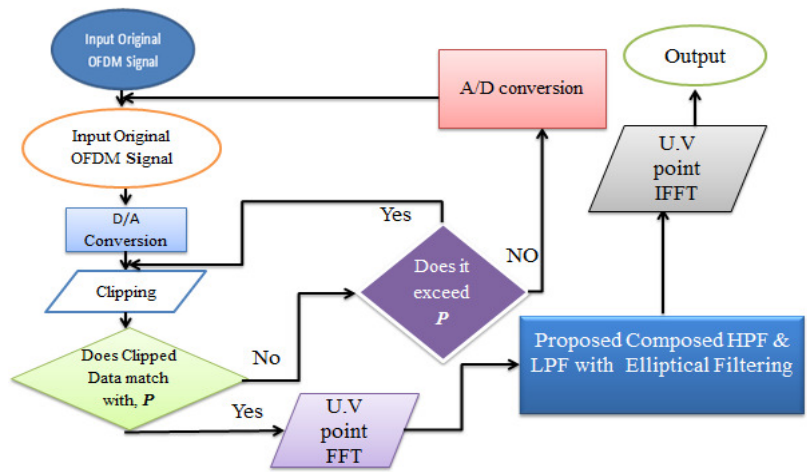

Figure 3. Proposed Algorithm for PAPR reduction

When ripple in the stopband approaches near or equal to zero, the filter becomes Chebyshev filter of type I. Again when ripple in the passband approaches near or equal to zero, the filter becomes Chebyshev filter of type II and when both ripple approach zero, this filter becomes a Butterworth filter.

The lowpass elliptic filter gain as a function of angular frequency ω is given by:

$$G_n(\omega) = \frac{1}{\sqrt{(1+\epsilon^2 R_n^2\left(\xi, \omega/\omega_0\right))}}$$

here $R_n$ is the *n*th-order elliptic rational function and $\omega_0$ is the cutoff frequency , $\epsilon$ is the ripple factor $\xi$ is the selectivity factor.

The ripple factor value specifies the passband ripple, while the combination of ripple factor and selectivity factor stipulate stopband ripple [12].

## 4. DESIGN PARAMETERS AND SIMULATION RESULTS

The observations were actually based on only QAM modulation. Table 1 shows the values of parameters used in the simulation for analysing the performance of amplitude clipping and filtering technique. It can be seen from the simulations results that it is possible for clipping and filtering scheme to reduce peak to average power ratio (PAPR). Simulation is done in the QAM modulation scheme i.e. 4-QAM has been used in OFDM generation which is very effective modulation techniques in 4G technologies having a bandwidth conserving modulation technique. The number of sub-carriers U is randomly having a sampling frequency of $F_S$= 8 MHz, satisfies the condition of orthogonality. PAPR(dB) of the original OFDM is computed by oversampling the number of sub-carriers K by the oversampling factor of L=8 while L=4 is enough, by insertion of (V-1)U zeros to reduce the ISI. Complementary Cumulative Distribution Function (CCDF) of PAPR is the measure of probability that how much higher is the PAPR value in comparison to PAPR (dB).

Figure 4 & 5 shows the PAPR distribution in case of N=128 & QPSK modulation for the existing[7] method & proposed method respectively.

Figure 5&7 shows the PAPR reduction for proposed filtering technique and figure 9&11 shows the BER performance.

Table 1. Parameter used for simulation

| Bandwidth, BW | 1 MHz |
|---|---|
| Over Sampling Factor, L | 8 |
| Sampling Frequency. $f_s$ | 8 MHz |
| Carrier Frequency, $f_c$ | 2 MHz |
| Cyclic Prefix Size | 32 |
| No.of Subcarrier/FFT size | 128 |
| Clipping Ratio | 0.8,0.8,1.0,1.2,1.4,1.6 |
| Modulation | QPSK/QAM |

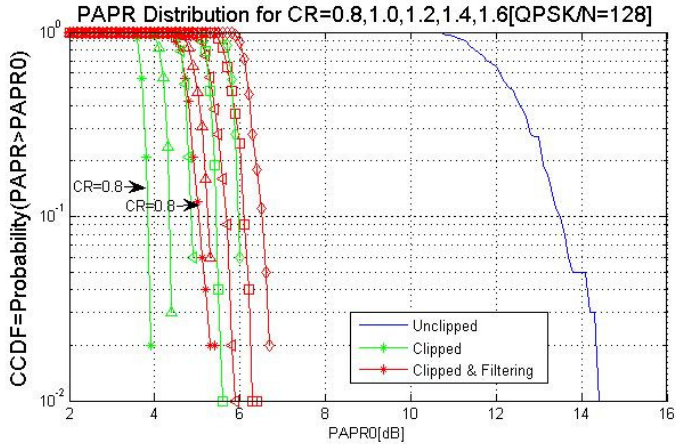

Figure 4. Existing method [7] of PAPR Reduction

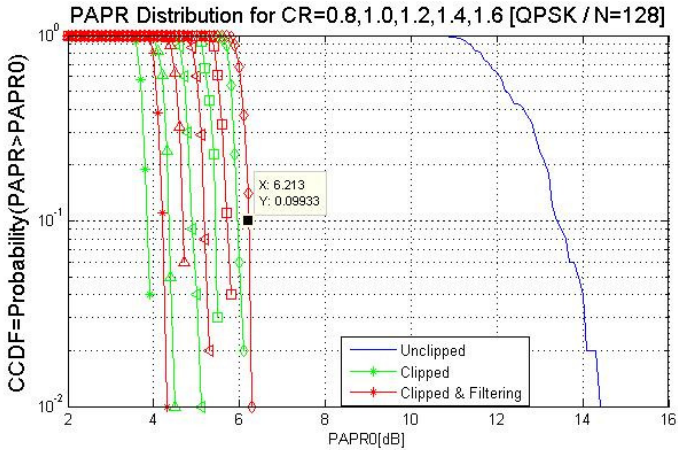

Figure 5. PAPR reduction by proposed method for QPSK

Table 2. Comparison of Existing method with Proposed Method for PAPR value [QPSK and N=128]

| CR value | PAPR value (dB) (Existing) | PAPR value (dB) (Proposed) | Improvement in PAPR value (dB) |
|---|---|---|---|
| Unclipped | 14.4 | 14.31 | 0.09 |
| 0.8 | 5.11 | 4.204 | 0.906 |
| 1.0 | 5.18 | 4.669 | 0.511 |
| 1.2 | 5.65 | 5.181 | 0.469 |
| 1.4 | 6.04 | 5.706 | 0.334 |
| 1.6 | 6.51 | 6.213 | 0.297 |

From Table 2, it is observed that, PAPR reduction improves in the proposed method.

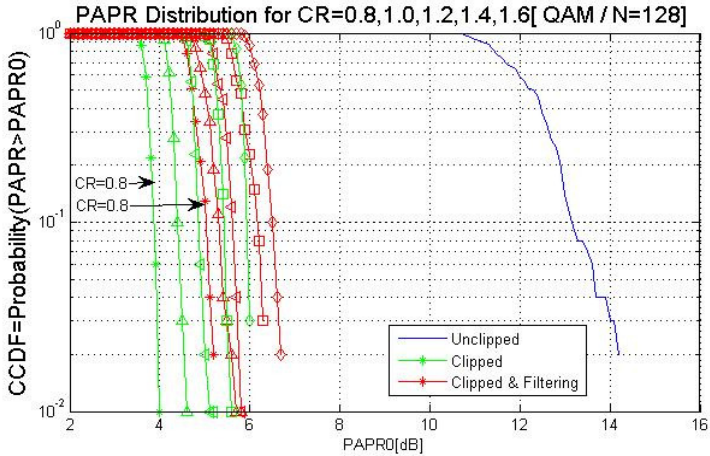

Figure 6. Existing method [7] of PAPR Reduction for QAM

Fig. 6 and Fig. 7 shows the PAPR distribution in case of N=128 and QAM modulation for the existing [7] method and proposed method respectively.

Table 3. Comparison of Existing method with Proposed Method for PAPR value [QAM and N=128]

| CR value | PAPR value (dB) (Existing) | PAPR value (dB) (Proposed) | Improvement in PAPR value (dB) |
|---|---|---|---|
| Unclipped | 14.11 | 14.24 | -0.13 |
| 0.8 | 4.97 | 4.199 | 0.771 |
| 1.0 | 5.25 | 4.655 | 0.595 |
| 1.2 | 5.67 | 5.201 | 0.469 |
| 1.4 | 6.09 | 5.717 | 0.373 |
| 1.6 | 6.51 | 6.254 | 0.256 |

From Table 3, it is observed that, PAPR reduction improves in the proposed method.

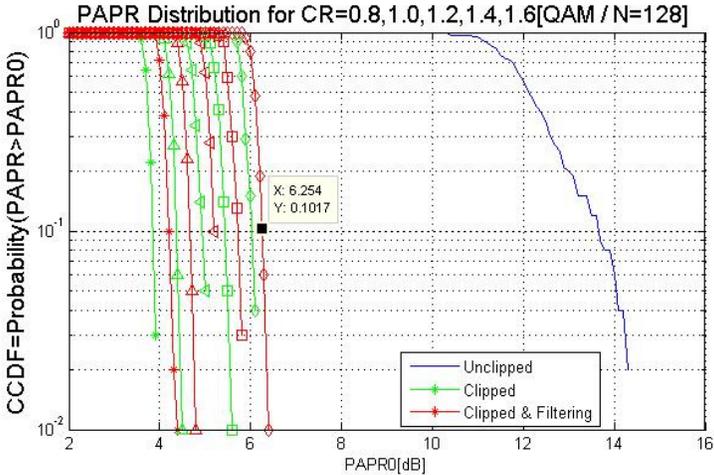

Figure 7. PAPR reduction by proposed method for QAM

Fig. 8 and Fig. 9 Shows the BER performance in case of N=128 and QPSK modulation for the existing [7] method and proposed method respectively.

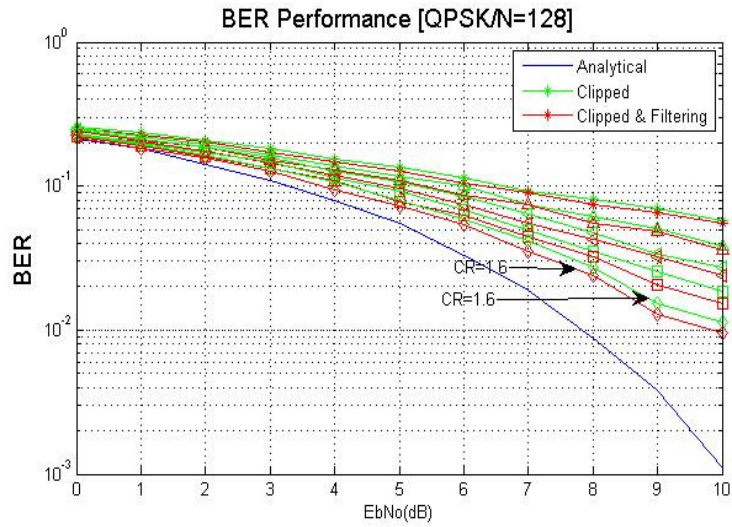

Figure 8. Existing method [7] of BER performance for QPSK

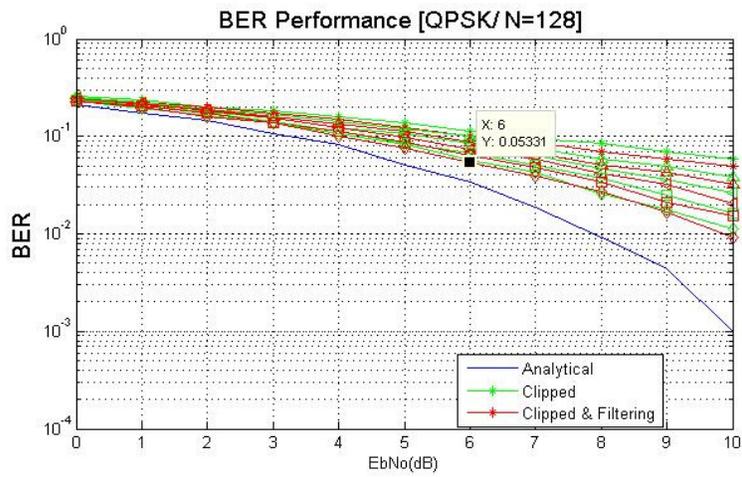

Figure 9. BER performance by Proposed Method for QPSK

Table 4. Comparison of BER performance for Existing & Proposed Method [QPSK and N=128]

| CR value | BER value (Existing) | BER Value (Proposed) | Difference in BER value |
|---|---|---|---|
| Analytical | 0.03281 | 0.03382 | -0.00101 |
| 0.8 | 0.07521 | 0.10131 | -0.0261 |
| 1.0 | 0.06163 | 0.08681 | -0.02518 |
| 1.2 | 0.04928 | 0.07358 | -0.02243 |
| 1.4 | 0.04025 | 0.06453 | -0.02428 |
| 1.6 | 0.03392 | 0.05331 | -0.01939 |

From Table 4, it is observed that, BER is increased a little bit compare to the existing method [7]. Fig. 10 and Fig. 11 Shows the BER performance in case of N=128 and QAM modulation for the existing [7] method and proposed method respectively.

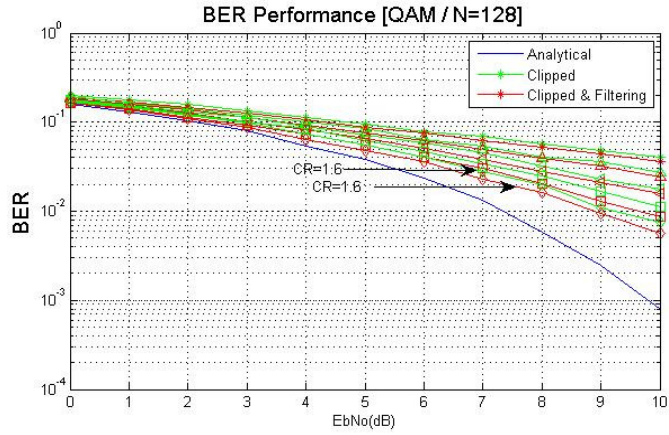

Figure 10. Existing method [7] of BER performance for QAM

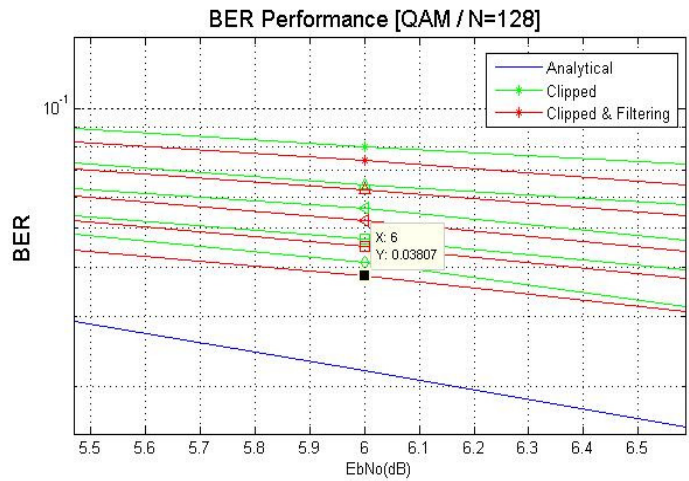

Figure 11. BER performance by Proposed Method for QAM

Table 5. Comparison of BER performance for Existing & Proposed Method [QAM and N=128]

| CR value | BER value (Existing) | BER Value (Proposed) | Difference in BER value |
|---|---|---|---|
| Analytical | 0.02335 | 0.02197 | 0.00138 |
| 0.8 | 0.07602 | 0.0739 | 0.00212 |
| 1.0 | 0.06256 | 0.0624 | 0.00016 |
| 1.2 | 0.05091 | 0.05213 | -0.00122 |
| 1.4 | 0.04089 | 0.04492 | -0.00403 |
| 1.6 | 0.03642 | 0.03807 | -0.00165 |

From Table 5, it is observed that, BER is increased a little bit compare to the existing method [7].

## 5. CONCLUSION AND FUTURE WORK

In this paper a comparative procedure of amplitude and clipping & filtering based PAPR reduction technique has been analysed with an existing method [7]. From the simulation we can see that PAPR reduces significantly compare to an existing one in the cost of little BER increase. We have simulated for QPSK and QAM modulation with 128 number of subcarrier respectively and executed it for the proposed method. It is observed that PAPR value is added with the increase of CR. In case of BER, with gradual increasing of CR value, the differences of BER value is decreasing. We can say QAM is ideal for higher order modulation. This work has been done under ideal channel condition. We will consider Rayleigh fading channel and new filtering techniques in our next works.

## REFERENCES

[1] R.W. Bauml, R.F.H. Fischer and J.B. Huber. "Reducing the peak-to-average power ratio of multicarrier modulation by selected mapping." Electronics Letters, 32(22):2056–7 ,1996/10/24.
[2] S.H. Han and J.H. Lee. "An overview of peak-to-average power ratio reduction Techniques for multicarrier transmission". IEEE Wireless Communications, 12(2):56–65,2005
[3] S.Y. L Goff, S.S. Al-Samahi, B.K. Khoo, C.C. Tsimenidis, and B.S. Sharif. "Selected mapping without side information for papr reduction in ofdm" IEEE Transactions on Wireless Communications, 8(7):3320 – 5/2009/07.
[4] P.Choudhury, A. Deshmukh." Comparison and analysis of PAPR reduction techniques in OFDM". IOSR Journal of Electronics and Communication Engineering (IOSR-JECE)ISSN: 2278-2834, ISBN: 2278-8735. Volume 4, Issue 5, (Jan - Feb. 2013), pp. 01-06.
[5] R. Ram, A. Hiradhar. "Efficient PAPR Reduction in OFDM Systems Based on a Companding Technique". International Journal of Science and Research (IJSR); Volume 3 Issue 7, July 2014; pp: 882-886
[6] M. Yang and Y. shin. "An adaptive tone injection for PAPR reduction", IEICE Electronics Express.vol:8,no:15; pp.1235-1239.
[7] M.M. Mowla. "Peak to average power ratio analysis and simulation in LTE system", M.Sc dissertation, Department of Electrical & Electronic Engineering, Rajshahi University of Engineering & Technology, Bangladesh, 2013.
[8] N. Arora & P. Singh. "Partial Transmit Sequence (PTS)-PAPR Reduction Technique in OFDM Systems with Reduced Complexity". Conference on Advances in Communication and Control Systems 2013. p.p. 355-359; 2013
[9] P.K. Sharma, "Power Efficiency Improvement in OFDM System using SLM with Adaptive Nonlinear Estimator" .World Applied Sciences Journal 7 (Special Issue of Computer & IT)., ISSN 1818.49520, pp.145-151, 2009.
[10] R.E. Regi, P.A. Haris. "Performance of PAPR Reduction in OFDM System with Complex Hadamard Sequence using SLM and Clipping". International Journal of Engineering and Advanced Technology (IJEAT).ISSN: 2249 – 8958, Volume-3, Issue-4, pp. 381-384, April 2014.
[11] Y.S. Cho, J. Kim, W.Y.Yang and C.G. Kang. "MIMO OFDM Wireless Communications with MATLAB", Singapore, John Wiley & Sons (Asia) Pt Ltd, 2010.
[12] Online : http://en.wikipedia.org/wiki/Elliptic_filter